%% file: KpKm_final.tex
\RequirePackage[modulo]{lineno}
\documentclass[aps,prl,twocolumn,superscriptaddress,groupedaddress]{revtex4}  % for review and submission
\usepackage{graphicx}  % needed for figures
\usepackage{dcolumn}   % needed for some tables
\usepackage{bm}        % for math
\usepackage{amssymb}   % for math
\usepackage{color}
\usepackage{amsmath}
\usepackage[percent]{overpic}

\def\Ecm {\ensuremath{\rm E_{\rm c.m.}}}
\def\Ebm {\ensuremath{\rm E_{\rm beam}}}

\def\epem {\ensuremath{e^+ e^-}}

\def\KK {\ensuremath{K^{+}K^{-}}}
\def\mevc {\ensuremath{\rm MeV/c}}

\def\mphi{ m_{\phi} = 1019.469 \pm 0.061~\rm MeV/c^2 }
\def\gammaphi{\Gamma_{\phi} = 4.249 \pm 0.015~\rm  MeV }
\def\gammaB{\Gamma_{\phi \to ee} B_{\phi \to K^+ K^-} = 0.670 \pm 0.022~\rm keV}
\def\gmu{a_{\mu}^{K^+K^-} = (19.33 \pm 0.40) \times 10^{-10}}
\def\ch2ndf{\chi^2/ndf = 25/20$ ($P(\chi^2)=20\%)}

\begin{document}

\widetext
\leftline{Primary authors: e.a.kozyrev@inp.nsk.su}
\leftline{To be submitted to PLB.}

\title{\bf{ \boldmath
Study of the process $\epem\to\KK$ in the center-of-mass energy range 1010--1060~MeV
with the CMD-3 detector 
% at \epem VEPP-2000 collider.
}}
\input author_list.tex

\date{\today}
\begin{abstract}
\hspace*{\parindent}
The process $\epem\to \KK$ has been studied
using $1.7 \times 10^6$ events from a data sample corresponding
to an integrated luminosity of  5.7 pb$^{-1}$
collected with the CMD-3 detector 
in the center-of-mass energy range 1010--1060 MeV. 
The cross section measured with an about 2\% systematic uncertainty and it is used 
to calculate the contribution to the anomalous magnetic moment of the muon 
$\gmu$, and to obtain the $\phi(1020)$ meson 
parameters.
We consider the relationship between the $\epem\to \KK$ and $\epem\to K^0_SK^0_L$ cross sections and compare it to the theoretical expectations.
\end{abstract}
\pacs{}
\maketitle
%\linenumbers
\setcounter{footnote}{0}
\section{Introduction}
\label{Introd}
Investigation of \epem~annihilation into hadrons at low energies provides 
unique information about interactions of light quarks. A precise measurement 
of the $\epem\to \KK$  cross section in the center-of-mass energy range 
\Ecm=1010--1060 MeV allows to obtain the $\phi(1020)$ meson parameters 
and to estimate a contribution of other light vector mesons, 
$\rho(770),~\omega(782)$, to the process studied. 
The $\epem\to\KK$ cross section, particularly in the $\phi$ meson energy region, 
is also required for a precise calculation of the hadronic contribution to the
muon anomalous magnetic moment, $a_\mu$, and the value of the running fine
structure constant at the $Z$ boson peak, 
$\alpha$(M$_Z$)~\cite{Davier_g_2}. 

The most precise cross section measurements performed by the 
CMD-2~\cite{cmdc} and BaBar~\cite{babarc} experiments have tension at
the level of more than 5\% (about 2.6 standard deviations) in the $\phi$ meson energy region.

Another motivation for the study arises from the comparison of the charged 
$\epem\to \KK$ and neutral $\epem\to K^0_SK^0_L$ final states. A significant 
deviation of the ratio of the coupling constants 
$\frac{g_{\phi \to K^{+}K^{-}}}{g_{\phi \to K^0_{S}K^0_{L}}}$ from a theoretical prediction
based on previous experiments (see the discussion in Ref.~\cite{Bramon}) 
requires a new precision measurement of the cross sections.

\section{CMD-3 detector and data set}
\label{CMD3}

The Cryogenic Magnetic Detector (CMD-3)  is a general purpose detector installed in one of the two 
interaction regions of the VEPP-2000 collider~\cite{vepp2000000} and is 
described elsewhere~\cite{cmd3}.
A detector tracking system consists of a cylindrical drift chamber (DC) 
and a double-layer cylindrical multiwire proportional chamber (Z-chamber), both installed 
inside a thin (0.2 $X_{0}$) superconducting solenoid with a 1.3 T field. 
The DC comprises of 1218 hexagonal cells and allows to measure charged particle 
momentum with a 1.5--4.5\% accuracy in the 100--1000~\mevc~ momentum range, it also
provides a measurement of the polar ($\theta$) and azimuthal ($\phi$) angles 
with a 20 mrad and 3.5-8.0 mrad accuracy, respectively. 
Amplitude information from the DC wires is used to measure the ionization 
losses $dE/dx$ of charged particles with a $ {\sigma}_{dE/dx}/<dE/dx>\approx$11--14\% 
accuracy for minimum ionization particles (m.i.p.). The Z-chamber with 
cathode strip readout is used to calibrate a DC longitudinal scale. 

An electromagnetic calorimeter comprised of a liquid xenon volume
of a 5.4 radiation length ($X_{0}$) thickness followed by CsI
crystals (8.1 $X_{0}$) outside of the solenoid in the barrel part and
BGO crystals (14.4 $X_{0}$) in the endcap parts~\cite{calorim1,calorim2}. A flux return yoke of the detector is surrounded by 
scintillation counters to veto cosmic events.

The beam energy \Ebm ~is monitored by using the 
back-scattering-laser-light system~\cite{compton,compton1}, which determines \Ecm~ at 
each energy point with an about 0.06 MeV systematic accuracy.  

Candidate events are recorded using signals from two independent trigger 
systems. One, a charged trigger, uses information only from DC cells 
indicating presence of at least one charged track, while another, a neutral 
trigger, requires an energy deposition in the calorimeter above \Ebm/2 or presence of more than 
two clusters above 25 MeV threshold.

To study a detector response for the investigated processes and to 
obtain a detection efficiency, we have developed a Monte Carlo (MC) 
simulation of the detector based on the GEANT4~\cite{GEANT4} package.
Simulated events are subject to all reconstruction and selection procedures. 
MC includes photon jet radiation by initial electron or positron (ISR) 
calculated according to Ref.~\cite{PJGen_sibid}. 

The measurement of the $\epem\to\KK$ cross section presented here is based 
on a  data sample collected at 24 energy points with a 5.7 pb$^{-1}$ 
integrated luminosity (IL) in the energy range \Ecm = 1010--1060 MeV
in 2012 and 2013.

\section{Event selection}
\label{Eventselect}
\begin{figure}
        \includegraphics[width=65mm]{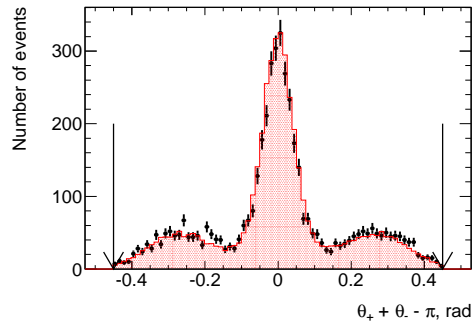}
                \caption{The polar collinearity $\theta_{K^+}+\theta_{K^-}-\pi$ 
for data (points) and MC (shaded histogram) at \Ebm = 530 MeV. 
        \label{deltath}}
\end{figure}

\begin{figure}
        \includegraphics[width=65mm]{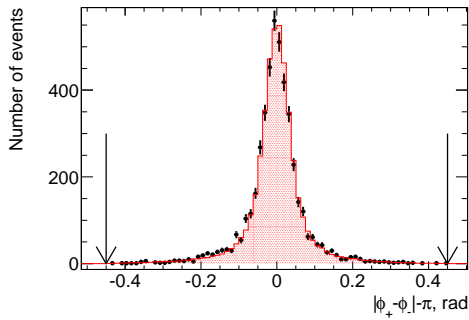}
                \caption{The azimuthal collinearity $|\phi_{K^+}-\phi_{K^-}|-\pi$ for data (points) and MC (shaded histogram) at \Ebm = 530 MeV. 
        \label{deltaphi}}
\end{figure}

\begin{figure} 
\begin{center}
   \begin{overpic}[scale=0.42]{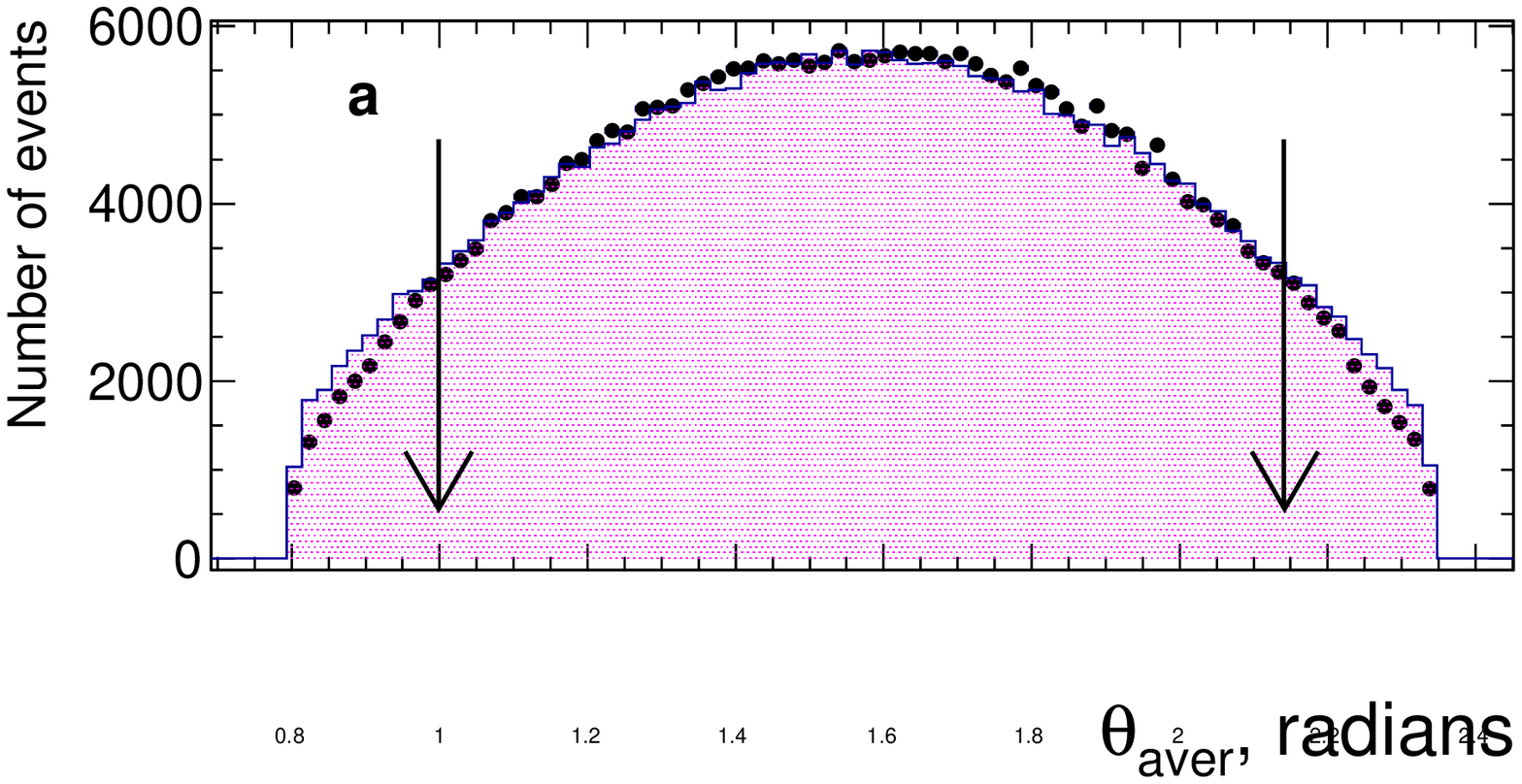}
 \put(0.2,0){\includegraphics[scale=0.42]{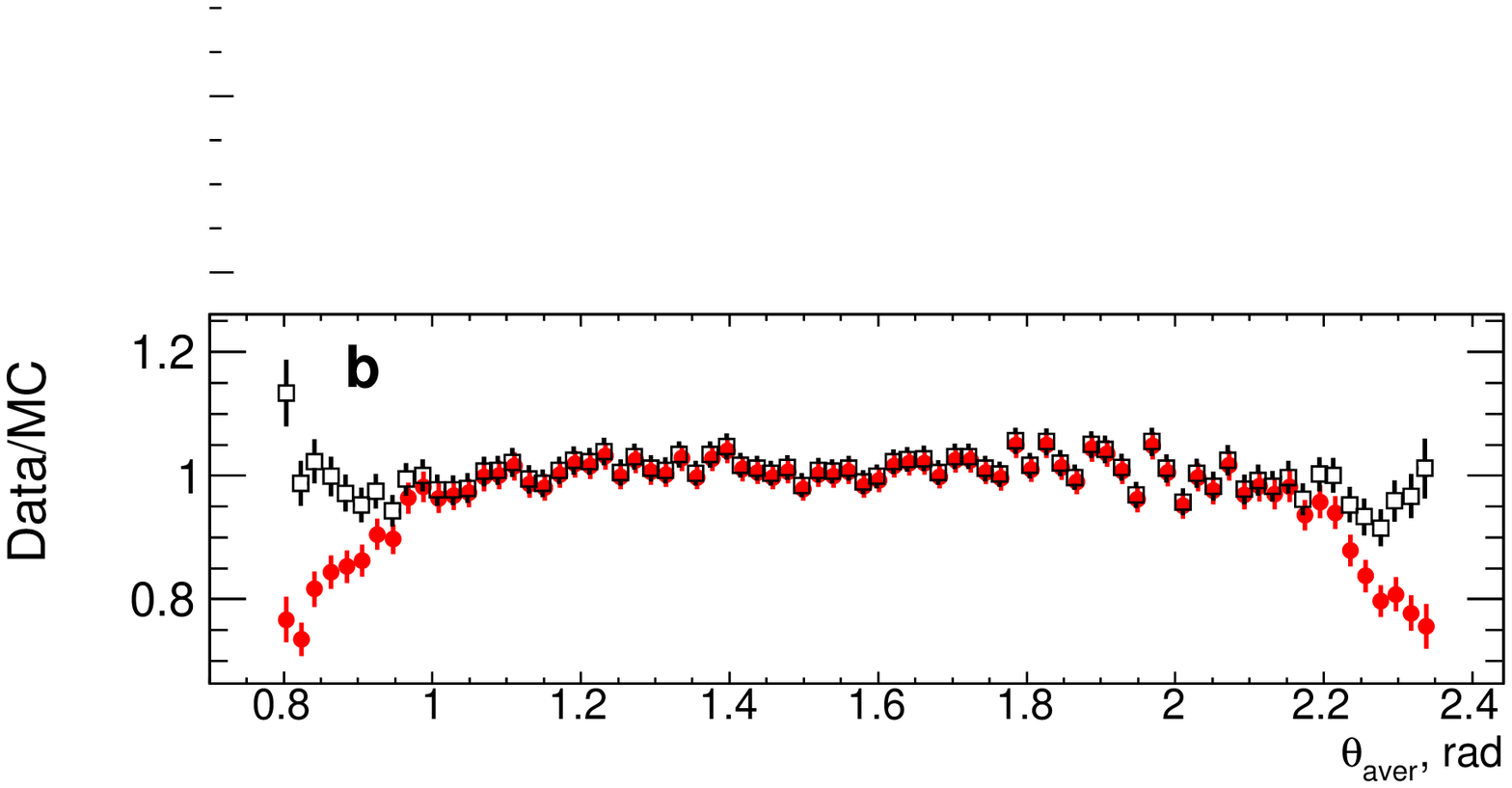}}
\end{overpic}
  \caption{(a) The average polar angle 
$\theta_{\rm aver}=(\theta_{K^{+}} + \pi - \theta_{K^{-}})/2 $ distribution for data 
(points) and MC (shaded) at \Ebm = 509.5 MeV. 
The MC histogram is normalized to six central bins of the data distribution. 
(b) The data-MC ratio before (points) and after (squares) applying efficiency 
corrections (see Sec.~\ref{syst}).
   \label{thaver}}
\end{center}
\end{figure}

\begin{figure}
        \includegraphics[width=87mm]{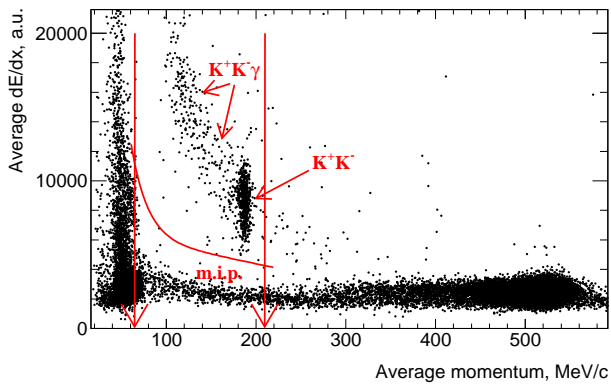}
                \caption{The ionization losses vs momentum for positive tracks 
for data at \Ebm = 530 MeV. The line shows selection of signal kaons. 
        \label{dedxp}}
\end{figure}

Selection of $\epem\to\KK$ candidates is based on the detection of two 
collinear tracks satisfying the following criteria: 

\begin{trivlist}
\item $\bullet$ The tracks originate from the beam interaction region within 
20 cm along the beam axis (Z-coordinate) and within 1 cm in the transverse 
direction.

\item $\bullet$ The polar and azimuthal collinearity are required to have 
$\Delta \theta = |\theta_{K^+}+\theta_{K^-}-\pi|,~\Delta \phi = ||\phi_{K^+}-\phi_{K^-}|-\pi| < $ 0.45 radians. 
The distributions of these parameters for data 
and MC at \Ebm~ = 530 MeV are shown in Figs.~\ref{deltath},\ref{deltaphi}, 
where the MC sample is normalized to data, and arrows demonstrate the applied 
requirement. Two additional bumps in the $\Delta \theta$ distribution 
are caused 
by a significant contribution of $K^+K^-\gamma$ events,  where $\gamma$ is 
emitted from the initial state (radiative return to the $\phi$ resonance).

\item $\bullet$ The tracks are required to have an average polar angle in the 
range 1 $< \theta_{\rm aver}=(\theta_{K^{+}} + \pi - \theta_{K^{-}})/2 < \pi$ - 1 radians.
The polar angle distribution is shown in Fig.~\ref{thaver} (a) where arrows 
demonstrate the applied restriction. Tracks out of the selected range do not 
pass all DC layers and are detected less efficiently (see the discussion 
in Sec.~\ref{syst}).
%Tracks, which do not belong to the chosen angle range, penetrate only part of DC layers and so are registered less effectively and rejected from the next analysis.

\item $\bullet$ Momenta of both tracks are required to be close to each 
other: $|p_1 - p_2|/|p_1 + p_2| <$ 0.3.

\item $\bullet$ The average momentum of two tracks is required to be in a 
range depending on \Ebm ~to minimize the background-to-signal ratio. An example 
of this restriction for \Ebm=530 MeV is shown in Fig.~\ref{dedxp} by arrows: 
the loss of signal events is less than 0.2\% according to MC. 

\item $\bullet$ In our energy range kaon ionization losses in DC are 
significantly larger than those for m.i.p.
due to a low momentum of kaons, $p$ = 100$\div$200 MeV/$c$. 
%due to a low velocity of kaons, $\beta_K$ = 0.2$\div$0.4. 
We require both tracks to have ionization 
losses above a value, which is calculated by taking into account the average 
value of dE/dx at the measured kaon momentum and dE/dx resolution. The line 
in Fig.~\ref{dedxp} shows an example of the applied selection. 
As seen in the figure, among selected events there are those with ISR photons,
which have smaller momentum and therefore larger dE/dx. 
Such events are also retained for the further analysis.
   
\end{trivlist}
 
\begin{figure}
        \includegraphics[width=80mm]{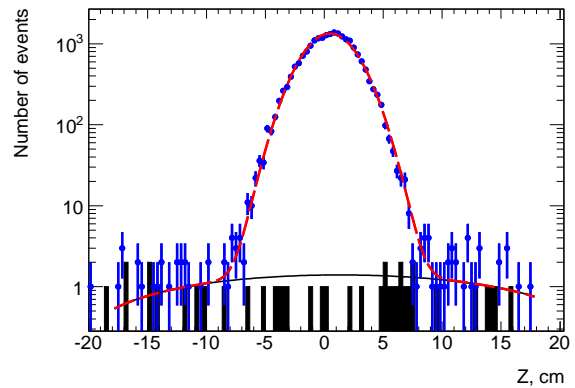}
                \caption{Approximation of the distribution of average 
Z-coordinates of selected tracks at \Ebm = 505 MeV. The long-dotted line 
corresponds to the signal, the solid line to the background. The shaded 
histogram shows the background distribution obtained using events 
at $E_{\rm c.m.}$ = 984 MeV.
        \label{fitZ}}
\end{figure}

The number of signal events is obtained using a fit of the 
average Z-coordinate distribution of two selected tracks with signal and 
background functions shown in Fig.~\ref{fitZ}. The shape of the signal function is 
described by a sum of two Gaussian distributions with parameters fixed from 
the simulation, and with additional Gaussian smearing to account for the 
difference in data-MC detector responses. For the background profile we use 
a second-order polynomial function, which describes well a distribution 
obtained at the energy \Ecm= 984 MeV below the threshold of the \KK~production
shown in Fig.~\ref{fitZ} by a shaded histogram. 
The level of background is estimated as less than 0.5\% for all energy points, 
except for the lowest energy  \Ecm= 1010.46 MeV, where the background is 
about 1.1\%.
The background is predominantly caused by the beam-gas interaction and 
interaction of particles lost from the beam at the vacuum pipe walls.
% of collider beams with vacuum pipe, producing electrons with spiral 
%trajectory transversely to beam line reconstructed in DC 
%as two collinear tracks.

As a result, we obtain (by fit) $1705060\pm 1306$ events of the 
process $\epem\to\KK$.


 \section{Detection efficiency}
\label{Eff}
The detection efficiency, $\epsilon_{\rm MC}$, is determined from MC by 
dividing the number of MC simulated events, after reconstruction and selection 
described above, to the total number of generated \KK~ pairs. 
The obtained $\epsilon_{\rm MC}$ is presented in Table~\ref{endtable}
and increases from 44\% to 55\%, and is primarily
determined by the restriction on the kaon polar angles and its decays in 
flight. Simulation 
of the ISR spectrum depends on the cross section under study and
this effect is taken into account by iterations. Influence of final-state 
radiation of real photons (FSR) on $\epsilon_{\rm MC}$ is examined by including 
into the  MC generator the FSR amplitude calculated according to scalar 
electrodynamics with pointlike K mesons~\cite{PJGen_sibid}. The observed 
change of $\epsilon_{\rm MC}$ is less than 0.1\%.   

Because of some data-MC inconsistency in the tracking efficiency, we introduce 
a correction equal to the ratio of a single-kaon-track efficiency in data and MC, 
$\epsilon^{+(-)}_{\rm EXP}/\epsilon^{+(-)}_{\rm MC}$. A detection efficiency 
corrected for detector effects is defined as
\begin{eqnarray}\label{eregggg}
\epsilon_{\rm det} = \epsilon_{\rm MC} \cdot \frac{\epsilon^{+}_{\rm EXP}}{\epsilon^{+}_{\rm MC}} \cdot \frac{\epsilon^{-}_{\rm EXP}}{\epsilon^{-}_{\rm MC}} ~,
\end{eqnarray}

\begin{figure} 
  \includegraphics[width=0.45\textwidth]{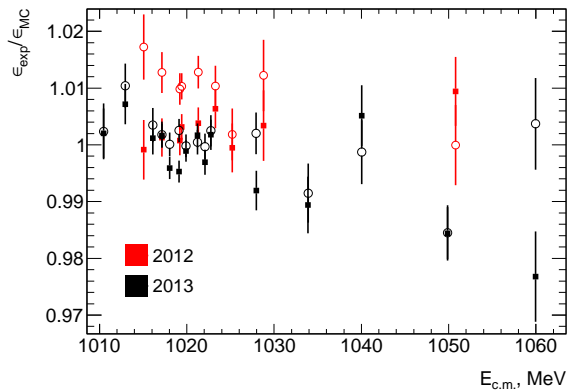}
  \caption{The EXP-MC ratio of the single-track efficiencies for positive 
$\frac{\epsilon^{+}_{\rm EXP}}{\epsilon^{+}_{\rm MC}}$ (squares) and negative kaons 
$\frac{\epsilon^{-}_{\rm EXP}}{\epsilon^{-}_{\rm MC}}$ (circles) for data collected 
in 2012 and 2013 runs.}
  \label{cor_p}
\end{figure}

The collinear configuration of the process and large ionization losses allow 
estimation of the single-kaon-track efficiency in data and MC to be performed 
by selecting a pure class of ``test" events with a detected positive or negative kaon, 
and checking how often we reconstruct the opposite track. The detection efficiencies for 
single positive and negative kaons increase from 80\% to 90\% in our energy range. 
The data-MC ratios $\frac{\epsilon^{+}_{\rm EXP}}{\epsilon^{+}_{\rm MC}}$ and 
$\frac{\epsilon^{-}_{\rm EXP}}{\epsilon^{-}_{\rm MC}}$ of the efficiencies are 
shown in Fig.~\ref{cor_p} for single positive (squares) and negative (circles) 
kaons vs c.m. energy,  and are used in Eq.~(\ref{eregggg}) to calculate 
the detection efficiency at each energy point.

\begin{figure} 
        \includegraphics[width=0.45\textwidth]{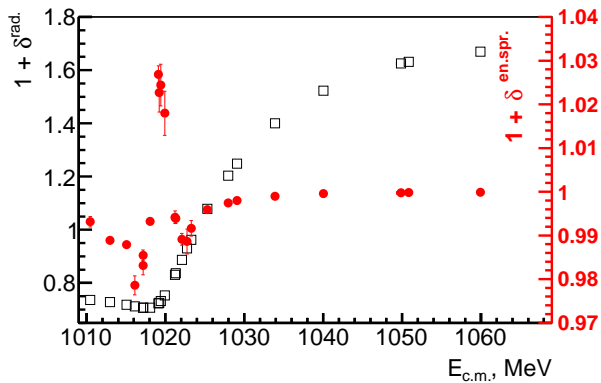}
        \caption{Radiative corrections $1+\delta^{\rm rad.}$ 
        (squares, left scale) and corrections $1+\delta^{\rm en.spr.}$ for 
the spread of collision energy (points, right scale).
                \label{radpfig}}    
\end{figure}

\begin{figure*} 
\begin{overpic}[scale=0.8]{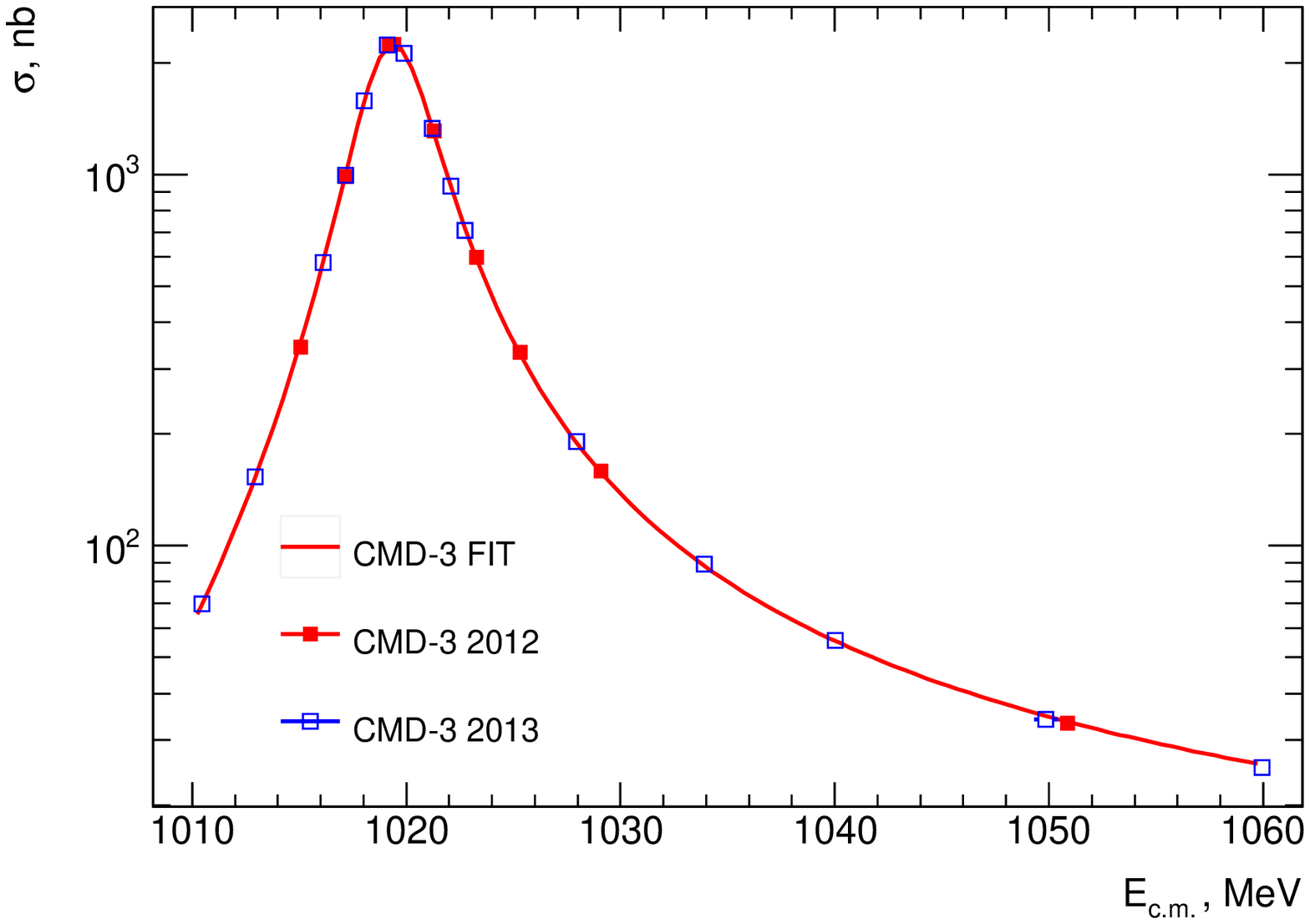} 
\put(49,32){\includegraphics[scale=0.36]{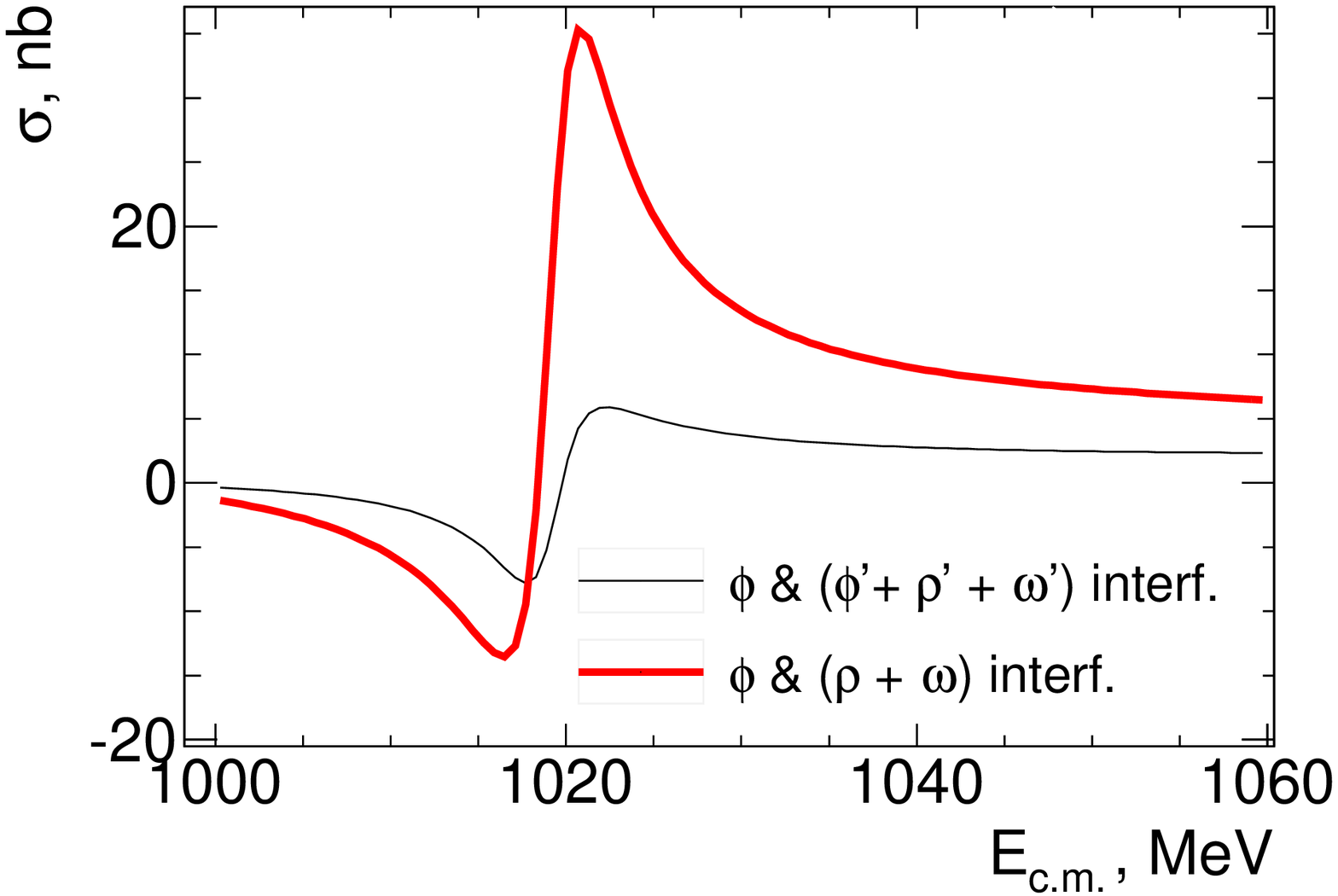}}
\end{overpic}
        \caption{\label{crosspict} Measured $e^+e^- \to K^+K^-$ cross section. 
The dots are experimental data, the curve is the fit described in the text.}    
\end{figure*}
 %(b) Relative difference between the data and fit. Comparison with other experimental data is shown. Statistical uncertainties only are included for data. The width of the band shows the systematic uncertainties in our experiment.

\begin{figure*} 
        \includegraphics[width=0.45\textwidth]{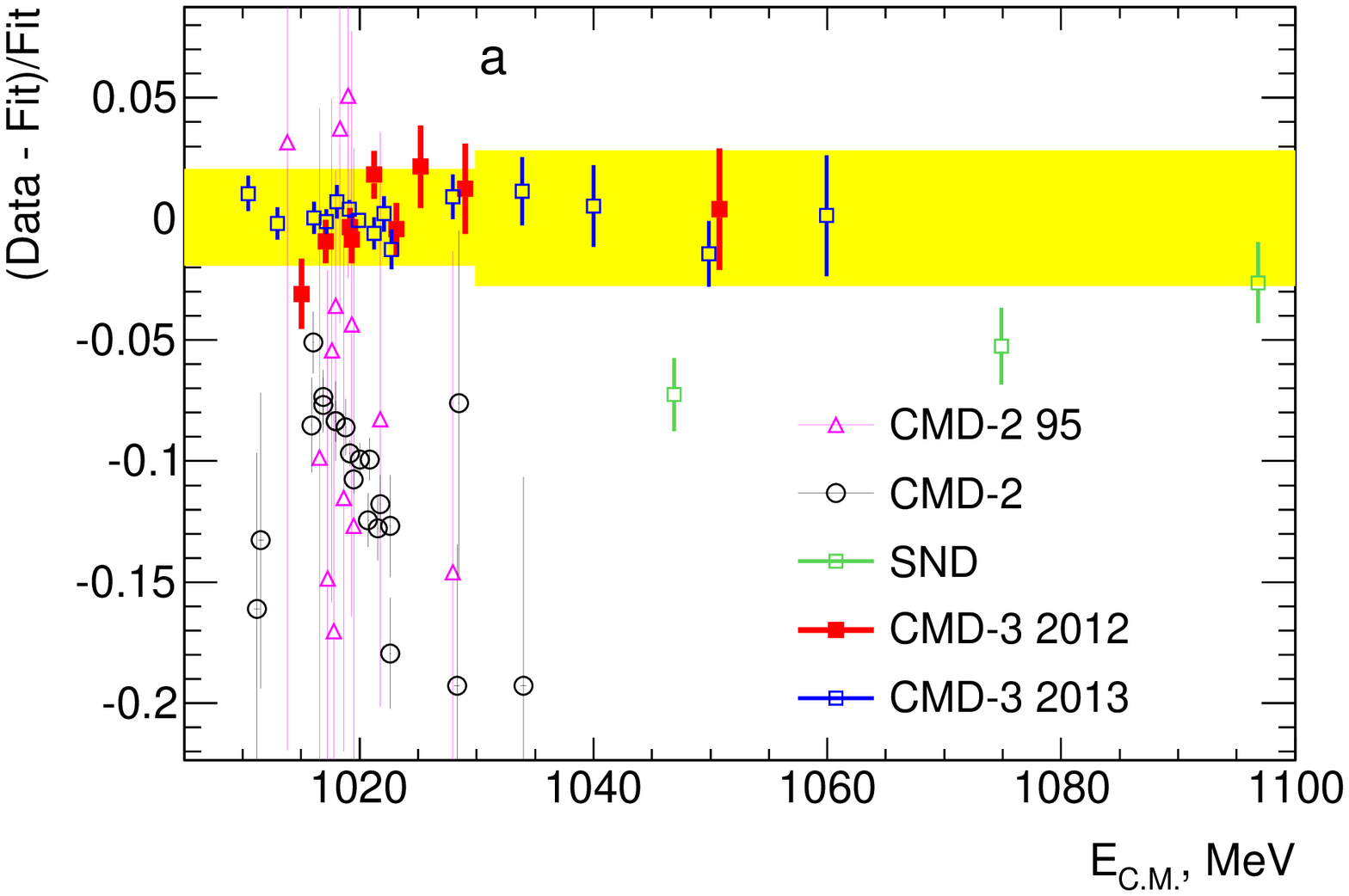}
        \includegraphics[width=0.45\textwidth]{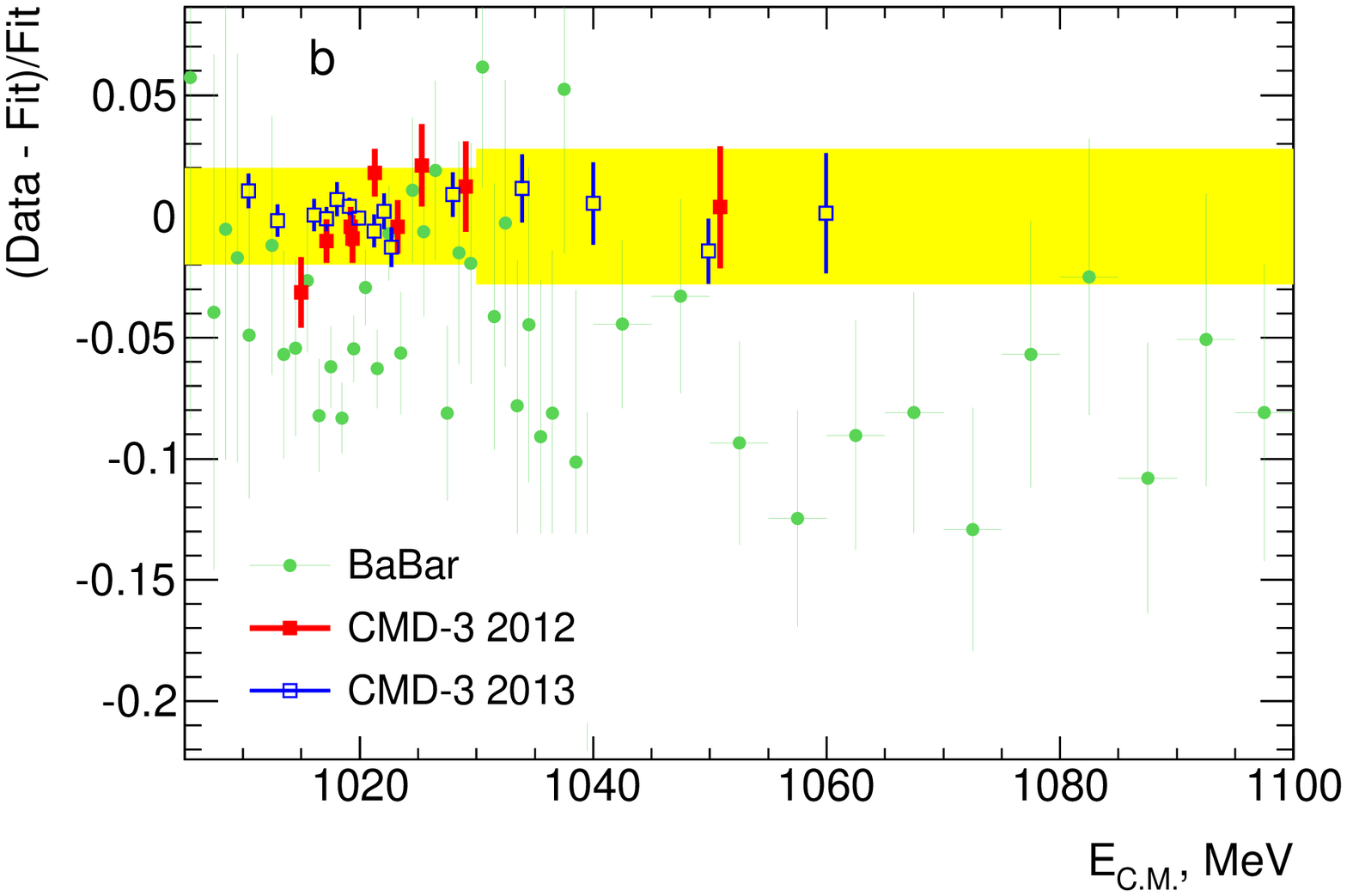}
        \caption{\label{crossrel} Comparison with other experimental data. 
Statistical uncertainties only are included for data. The width of the band 
shows the systematic uncertainties in this study.}    
\end{figure*}
\section{Cross section of $e^+ e^- \to K^{+} K^{-}$}
The experimental Born cross section of the process $e^+ e^- \to K^+ K^-$  
has been calculated for each energy point according to the expression: 
\begin{eqnarray} \label{crosss} 
    \sigma^{\rm born} &=& \frac{N_{\rm exp}}{\epsilon_{\rm det}\cdot\epsilon_{\rm trig} \cdot \rm IL \cdot  (1 + \delta^{\rm rad.})}\cdot (1+\delta^{\rm en.spr.}),  
\end{eqnarray}
where
%$\epsilon_{\rm reg}$ is the registration efficiency, 
$\epsilon_{\rm trig}$ is a trigger efficiency,
$\rm IL$ is the integrated luminosity, 
$1+\delta^{\rm en.spr.}$ represents a correction due to the energy spread 
of the electron-positron beams, and  $1+\delta^{\rm rad.}$ is an 
initial-state radiative correction. 
The integrated luminosity $\rm IL$ is determined by the processes 
$e^+e^- \to e^+e^-$ and $e^+e^- \to \gamma\gamma$ with an about 1\%~\cite{lum,lum1} 
accuracy. The correction $1+\delta^{\rm rad.}$, shown 
by squares in Fig.~\ref{radpfig}, is calculated using the radiative 
structure function, known with an accuracy better than 0.1\%~\cite{radcorFadin}.

The electron-positron c.m. energy spread, $\sigma_{\rm E_{c.m.}}$, typically 
about 300 keV, changes the visible cross section, and to take into account 
this effect we apply the following correction:

\begin{eqnarray}\label{spread_correction} 
1+\delta^{\rm en.spr.}(\rm E_{c.m.}) = \frac{1}{\sqrt{2\pi}\sigma_{\rm E_{c.m.}} }\cdot ~~~~~~~~~~~~~~~~~~~\\ \nonumber
\cdot\frac{\int \rm dE'_{\rm c.m.} \sigma^{\rm born}(\rm E'_{c.m.}) (1 + \delta^{\rm rad.}(\rm E'_{c.m.}))  e^{-\frac{(\rm E'_{c.m.} - \rm E_{c.m.})^2}{2\sigma^2_{\rm E_{c.m.}}} }} {\sigma^{\rm born}(\rm E_{c.m.}) (1 + \delta^{\rm rad.}(\rm E_{c.m.}))},
\end{eqnarray}
which depends on the cross section $\sigma^{\rm born}$, 
radiative correction $(1 + \delta^{\rm rad.})$, 
and is calculated by iterations in the same way as $\epsilon_{\rm MC}$ and 
$(1 + \delta^{\rm rad.})$. The calculated ($1+\delta^{\rm en.spr.}$) value for 
each energy point is shown in Fig.~\ref{radpfig} by circles (right scale), 
and has the maximum value of 1.026$\pm$0.006 at the 
peak of the $\phi$ resonance.

The trigger efficiency,  $\epsilon_{\rm trig}$, is studied using responses of 
two independent triggers, charged and neutral, for selected signal events, 
and is found to be close to unity $\epsilon_{\rm trig} = 0.998 \pm 0.001$ for 
applied selections.

The resulting cross section is listed in Table~\ref{endtable} at each energy 
point and shown in Fig.~\ref{crosspict}. The statistical error includes 
fluctuations of signal and Bhabha events, used for the luminosity calculation, 
and fluctuations of the uncertainty on the c.m. energy measurement,  
$\delta E_{\rm c.m.}$, calculated as $|\frac{\partial\sigma^{\rm born}}{\partial E_{\rm c.m.}}| \times \delta E_{\rm c.m.}$. 

\begin{table*}
\caption{The c.m. energy $E_{\rm c.m.}$, number of selected signal events $N$,  
uncorrected and corrected detection efficiencies $\epsilon_{\rm MC}$ and $\epsilon_{\rm det}$,
radiative correction factor 1 + $\delta^{\rm rad.}$,
correction for the spread of collision energy $ 1 + \delta^{\rm en.spr.}$,
integrated luminosity $\rm IL$, 
and Born cross section $\sigma$ of the process $e^+ e^- \to K^{+}K^{-}$ where only statistical errors are presented.
}
\label{endtable}
\begin{center}
\begin{tabular}[t]{cccccccc}
\hline
   $\Ecm$, MeV     &    $N$ events    &   $\epsilon_{\rm MC}$   &  $\epsilon_{\rm det}$   & 1 + $\delta^{\rm rad.}$  &     $ 1 + \delta^{\rm en.spr.}$  &  $\rm IL$, nb$^{-1}$ &   $\sigma$, nb \\
\hline
 1010.47 $\pm$ 0.01 & 21351 $\pm$ 145 & 0.439 & 0.441 & 0.735 & 0.993 & 936.05 $\pm$ 1.44 & 69.87 $\pm$ 0.50 \\ 
 1012.96 $\pm$ 0.01 & 26882 $\pm$ 164 & 0.485 & 0.493 & 0.728 & 0.988 & 485.36 $\pm$ 1.04 & 152.45 $\pm$ 1.01 \\ 
 1015.07 $\pm$ 0.02 & 6031 $\pm$ 78 & 0.502 & 0.510 & 0.718 & 0.987 & 47.91 $\pm$ 0.33 & 341.10 $\pm$ 5.11 \\ 
 1016.11 $\pm$ 0.01 & 41260 $\pm$ 201 & 0.510 & 0.513 & 0.712 & 0.978 & 192.11 $\pm$ 0.66 & 575.08 $\pm$ 3.84 \\ 
 1017.15 $\pm$ 0.02 & 176768 $\pm$ 421 & 0.515 & 0.517 & 0.706 & 0.983 & 478.99 $\pm$ 1.04 & 993.19 $\pm$ 5.02 \\ 
 1017.16 $\pm$ 0.02 & 22243 $\pm$ 149 & 0.517 & 0.524 & 0.706 & 0.985 & 60.15 $\pm$ 0.30 & 984.71 $\pm$ 8.89 \\ 
 1018.05 $\pm$ 0.03 & 279733 $\pm$ 529 & 0.521 & 0.519 & 0.706 & 0.993 & 478.34 $\pm$ 1.04 & 1584.27 $\pm$ 11.00 \\ 
 1019.12 $\pm$ 0.02 & 270045 $\pm$ 520 & 0.525 & 0.524 & 0.721 & 1.026 & 328.62 $\pm$ 0.86 & 2228.59 $\pm$ 8.13 \\ 
 1019.21 $\pm$ 0.03 & 44051 $\pm$ 209 & 0.525 & 0.531 & 0.724 & 1.022 & 52.75 $\pm$ 0.34 & 2230.81 $\pm$ 18.14 \\ 
 1019.40 $\pm$ 0.04 & 30539 $\pm$ 174 & 0.526 & 0.533 & 0.730 & 1.024 & 36.05 $\pm$ 0.29 & 2233.66 $\pm$ 22.07 \\ 
 1019.90 $\pm$ 0.02 & 391083 $\pm$ 626 & 0.527 & 0.527 & 0.752 & 1.017 & 472.34 $\pm$ 1.04 & 2127.07 $\pm$ 6.46 \\ 
 1021.22 $\pm$ 0.03 & 134598 $\pm$ 365 & 0.532 & 0.533 & 0.829 & 0.994 & 228.34 $\pm$ 0.72 & 1325.01 $\pm$ 9.01 \\ 
 1021.31 $\pm$ 0.01 & 27717 $\pm$ 165 & 0.531 & 0.540 & 0.835 & 0.993 & 46.85 $\pm$ 0.33 & 1308.31 $\pm$ 12.50 \\ 
 1022.08 $\pm$ 0.03 & 89487 $\pm$ 299 & 0.532 & 0.530 & 0.885 & 0.989 & 201.62 $\pm$ 0.68 & 933.95 $\pm$ 6.81 \\ 
 1022.74 $\pm$ 0.03 & 41756 $\pm$ 204 & 0.534 & 0.536 & 0.928 & 0.988 & 116.71 $\pm$ 0.52 & 710.23 $\pm$ 5.86 \\ 
 1023.26 $\pm$ 0.04 & 19718 $\pm$ 140 & 0.536 & 0.545 & 0.961 & 0.991 & 62.91 $\pm$ 0.38 & 595.03 $\pm$ 6.56 \\ 
 1025.32 $\pm$ 0.04 & 7023 $\pm$ 84 & 0.537 & 0.538 & 1.077 & 0.995 & 36.32 $\pm$ 0.29 & 334.77 $\pm$ 5.55 \\ 
 1027.96 $\pm$ 0.02 & 24236 $\pm$ 156 & 0.540 & 0.536 & 1.200 & 0.997 & 195.83 $\pm$ 0.67 & 191.64 $\pm$ 1.74 \\ 
 1029.09 $\pm$ 0.02 & 5786 $\pm$ 76 & 0.542 & 0.550 & 1.244 & 0.997 & 52.94 $\pm$ 0.35 & 159.94 $\pm$ 2.95 \\ 
 1033.91 $\pm$ 0.02 & 11752 $\pm$ 108 & 0.546 & 0.535 & 1.392 & 0.998 & 175.55 $\pm$ 0.64 & 89.65 $\pm$ 1.24 \\ 
 1040.03 $\pm$ 0.05 & 9143 $\pm$ 95 & 0.551 & 0.553 & 1.509 & 0.999 & 195.91 $\pm$ 0.68 & 55.87 $\pm$ 0.94 \\ 
 1049.86 $\pm$ 0.02 & 14818 $\pm$ 122 & 0.553 & 0.536 & 1.604 & 0.999 & 499.59 $\pm$ 1.09 & 34.47 $\pm$ 0.47 \\ 
 1050.86 $\pm$ 0.04 & 4441 $\pm$ 67 & 0.554 & 0.559 & 1.609 & 0.999 & 146.31 $\pm$ 0.59 & 33.89 $\pm$ 0.84 \\ 
 1059.95 $\pm$ 0.02 & 4594 $\pm$ 68 & 0.553 & 0.543 & 1.640 & 0.999 & 198.86 $\pm$ 0.69 & 25.93 $\pm$ 0.64 \\ 
 
\hline
\end{tabular}
\end{center}
\end{table*}

\section{Systematic uncertainties}
\label{syst}
The uncertainty on the $e^+ e^- \to K^{+}K^{-}$ cross section is 
dominated by the accuracy of the detection efficiency $\epsilon_{\rm det}$ 
calculation.  

The systematic uncertainty of the data-MC ratios in Eq.~(\ref{eregggg}) is 
estimated by applying  different selection requirements on the ``test'' events 
and does not exceed 1\%, however, for five energy points with $\Ecm > 1030$ 
MeV we increase the uncertainty to 2\%. 

The data-MC difference in the polar angle distributions of kaons is shown in 
Fig.~\ref{thaver}(b) by circles. The observed difference is
due to incorrect simulation of detector resolution, angular dependence of the 
track reconstruction and trigger efficiency, and  uncertainty on the 
calibration of the DC longitudinal scale. 
We tune our simulation to match the detector angular and momentum resolutions 
(see Fig.~\ref{deltath}), study  angular dependence of the track reconstruction
efficiency using a single-track ``test'' sample, and study response of two 
independent triggers as a function of the track polar angle. The data-MC ratio of the polar 
angle distributions after applied corrections is shown in 
Fig.~\ref{thaver}(b) by squares.

To estimate influence of the remaining angular uncertainty on the measured 
cross section we divide all data into three independent samples with 
$\theta \in$ [0.95 : 1.35], [1.35 : $\pi-$1.35] and [$\pi-$1.35 : $\pi-$0.95] 
radians. By separately calculating all parameters in Eq.~(\ref{crosss}) for 
three regions and comparing the obtained cross sections we estimate the 
corresponding uncertainty as the average difference of the samples
to be 1\%. 

To check the quality of the DC chamber scale calibration we extrapolate the 
reconstructed kaon tracks from DC to ZC and compare it with the position of 
the ZC response: a possible sytematic uncertainty is less than 0.3\%. 

The total systematic uncertainty in the reconstruction efficiency is estimated 
as 1.6\%, but increased to 2.5\% for five energy points for \Ecm$>$1030 MeV.   

To estimate uncertainty on the background subtraction procedure
we use the data accumulated at the energy point \Ecm= 984 MeV below the 
reaction threshold. Applying our selection criteria we obtain the number of 
background events, $\rm N_{984}$, and then estimate the number of background 
events for each energy point using the integrated luminosities IL(s) as:
\begin{eqnarray} 
\rm N_{\rm bkg}(\Ecm) = \rm N_{984} \cdot \frac{\rm IL(\Ecm)}{\rm IL(984)}.
\end{eqnarray} 
The difference in the calculated number of background events and events 
obtained by the approximation of the Z-coordinate distribution (see 
Sec.~\ref{Eventselect}) gives  less than 0.3\% uncertainty of the cross 
section: this value is used as an estimate of the corresponding systematic 
uncertainty.

A significant part of selected signal events include ISR photons, which 
should be taken into account in $\epsilon_{\rm det}$ and  $1 + \delta^{\rm rad.}$. 
The photon spectrum is calculated by a convolution of the radiator 
function~\cite{radcorFadin} and Born cross section $\sigma^{\rm born}$(\Ecm) 
which is known with uncertainties discussed above. By varying $N_{\rm exp}$, IL,
\Ecm, and $\epsilon_{\rm det}$ in Eq.~(\ref{crosss}) according to their 
uncertainties and repeating the calculation of $\sigma^{\rm born}$(\Ecm) and 
$1 + \delta^{\rm rad.}$ we estimate the uncertainty on the cross section as 
0.1\% (0.8\% for energy points with $\Ecm > 1030$ MeV). These values are 
quadratically summed with the 0.1\% theoretical uncertainty of the radiator 
function.

The systematic uncertainties contributing to the measured cross section are 
listed in Table~\ref{endtable_syst}, and the quadratic sum gives  
2.0\% (2.8\% for $\Ecm > 1030$ MeV) as the total systematic uncertainty. 
\begin{table}
\begin{center}
\caption{Summary of systematic uncertainties on the 
$e^+e^- \to K^{+} K^{-}$ cross section measurement}
\label{endtable_syst}
\begin{tabular}[t]{lc}
\hline
Source                        &   Uncertainty, \% \\
\hline
Signal selection               & 0.3       \\
Detection efficiency          & 1.6(2.5)       \\
Radiative correction          & 0.15(0.80)      \\
Energy spread correction      & 0.3       \\
Trigger efficiency            & 0.1       \\
Luminosity                    & 1.0       \\
\hline
Total                         & 2.0(2.8)        \\
\hline
\end{tabular}
\end{center}
\end{table}

\section{Approximation of the $e^+ e^- \to K^{+}K^{-}$ cross section}
\label{lapproxim}
The measured cross section defined by 
Eq.~(\ref{crosss}) includes a vacuum polarization factor, Coulomb interaction 
between $K^{+} K^{-}$, and final-state radiation of real photons $\gamma_{FSR}$. 
We approximate the energy dependence  
of the cross section according to the vector meson dominance (VMD) model as a 
squared sum of the $\rho,~\omega,~\phi$-like amplitudes~\cite{Kuhn}:
\begin{eqnarray} 
\sigma(s) \equiv \sigma_{e^+ e^- \to K^{+} K^{-}}(s) = \frac{8 \pi \alpha}{3 s^{5/2}} p_{K}^{3} \frac{Z(s)}{Z(m_{\phi}^2)} \Biggr| \frac{g_{\phi \gamma} g_{\phi KK}}{D_{\phi}(s)} \nonumber \\ 
%e^{i\psi_{\phi}}
+ r_{\rho,\omega} \times [\frac{g_{\rho \gamma} g_{\rho KK}}{D_{\rho}(s)} 
+ 
\frac{g_{\omega \gamma} g_{\omega KK}}{D_{\omega}(s)}] + A_{\phi', \rho', \omega'}\Biggr|^{2},
\label{ksklxs}
\end{eqnarray}
where $s = E_{\rm c.m.}^2$, $p_{K}$ is a kaon momentum, 
\begin{eqnarray} \label{FGSZ}
Z(s) =  \frac{\pi\alpha/\beta}{1-exp(-\pi\alpha/\beta)}  \Biggr(1+\frac{\alpha^2}{4\beta^2}\Biggr)
\end{eqnarray}
is the Sommerfeld-Gamov-Sakharov factor that can be obtained by solving 
the Schr$\ddot{\rm o}$dinger equation in a Coulomb potential for a P-wave 
final state with velocity $\beta = \sqrt{1-4m^2_{K}/s}$, $D_{V}(s)$ is the 
inverse propagator of the vector state V: 
 \begin{eqnarray}
 \label{propogator}
 D_{V}(s) = m_{V}^{2} -s - i \sqrt{s}\Gamma_{V}(s).
 \end{eqnarray}
Here $m_{V}$ and $\Gamma_{V} $ are mass and width of the major intermediate 
resonances: $V = \rho(770),~ \omega(782),~ \phi(1020)$. 
For the energy dependence of the  $\phi$ meson width we use 
\begin{eqnarray*}
\Gamma_{\phi}(s) = \Gamma_{\phi} \cdot \Biggr( B_{K^+K^-}\frac{m_{\phi}^2F_{K^+K^-}(s)}{sF_{K^+K^-}(m_{\phi}^2)}  \\ \nonumber
+B_{K^0_SK^0_L}\frac{m_{\phi}^2F_{K^0_SK^0_L}(s)}{sF_{K^0_SK^0_L}(m_{\phi}^2)} +
B_{\pi^+\pi^-\pi^0}\frac{\sqrt{s}F_{\pi^+\pi^-\pi^0}(s)}{m_{\phi}F_{\pi^+\pi^-\pi^0}(m_{\phi}^2)}  \\ \nonumber
+B_{\eta\gamma}\frac{F_{\eta\gamma}(s)}{F_{\eta\gamma}(m_{\phi}^2)} \Biggr),
\end{eqnarray*}
where $F_{K\bar{K}} = (s/4 - m_{K}^2)^{3/2},~F_{\eta\gamma}(s) = (\sqrt{s}(1-m_{\eta}^2/s))^3$, 
and for the $F_{\pi^+\pi^-\pi^0}(s)$ calculation the model assuming the 
$\phi \to \rho\pi \to \pi^+\pi^-\pi^0$ decay is used~\cite{NNAchasov}. The magnitudes of 
$\Gamma_{\rho}(s)$  and $\Gamma_{\omega}(s)$ are calculated in the same way using 
the corresponding branching fractions~\cite{PDG}.
The coupling constants of the intermediate vector meson $V$ with initial and 
final states can be presented as: 
\begin{eqnarray*}
g_{V \gamma} = \sqrt{\frac{3 m_{V}^{3} \Gamma_{Vee}}{4 \pi \alpha}};~g_{V K^{+}K^{-}} = \sqrt{\frac{6 \pi m_{V}^{2} \Gamma_{V} B_{VK^+K^-}}{p^{3}_{K}(m_{V})}},
\end{eqnarray*}
where $\Gamma_{Vee}$ and $B_{VK^{+}K^{-}}$ are electronic width and decay 
branching fraction to a kaon pair. In our approximation we use the table 
values of mass, total width, and electronic width of the $\rho(770)$ and 
$\omega(782)$:  $\Gamma_{\rho\to ee} = 7.04\pm0.06~\rm keV,~ \Gamma_{\omega\to ee} = 0.60\pm0.02~\rm keV$~\cite{PDG}.
For the {\it a priori} unknown couplings of the $\rho(770)$ and $\omega(782)$ 
to the pair of kaons we use the relation 
\begin{eqnarray} 
\label{sfdsfsdfsd}
g_{\omega K^+K^-} = g_{\rho K^+K^-} = - g_{\phi K^+K^-}/\sqrt{2},
\end{eqnarray}
based on the quark model with ``ideal" mixing and exact SU(3) symmetry of 
u-, d-, and s-quarks~\cite{Kuhn}. In order to take into account a possible 
breaking of these assumptions, both $g_{\rho K^+K^-}$ and $g_{\omega K^+K^-}$ are 
multiplied by the common complex constant $r_{\rho/\omega}$.  

The amplitude  $A_{\phi', \rho', \omega'}$ denotes a contribution of  the higher
vector mesons $\omega(1420),~\rho(1450)$, $\omega(1650)$, $\phi(1680)$ and 
$\rho(1700)$ to the 
$\phi(1020)$ mass region. Using BaBar~\cite{babarc} and SND~\cite{SND2016} 
data above $\sqrt{s}=$ 1.06 GeV for the process $e^+e^- \to K^{+}K^{-}$ we 
extract a contribution of these states.

We perform a fit to the $e^+e^- \to K^{+}K^{-}$ cross section with floating 
$m_{\phi},~\Gamma_{\phi}$, $\Gamma_{\phi \to ee}\times B_{\phi \to K^{+}K^{-}}$ (or 
alternatively $B_{\phi \to ee}\times B_{\phi \to K^{+}K^{-}}$) and $r_{\rho/\omega}$ 
parameters: the fit yields $\ch2ndf$. The fit 
result is shown in Fig.~\ref{crosspict}. Figures~\ref{crossrel} show the 
relative difference between the obtained data and the fit curve. Only 
statistical errors are shown and the width of the band corresponds to the 
systematic uncertainty of the cross section.
In Fig.~\ref{crossrel} (a) we compare our result with the 
previous Novosibirsk measurements~\cite{cmdc,cmd_old,SND2016} while 
Fig.~\ref{crossrel} (b) shows a comparison with the recent BaBar 
experiment~\cite{babarc}. The obtained parameters of the $\phi$ meson in 
comparison with the values of other measurements are presented in 
Table~\ref{endtable_phi_parameters}. The first uncertainties are statistical 
and the second are systematic, resulting from the 60 keV accuracy in the \Ecm ~
measurements and errors listed in Table~\ref{endtable_syst}. From the fit 
we obtain $\rm Re~(r_{\rho/\omega}) = 0.95 \pm 0.03$ while an imaginary part is 
compatible with zero.
The  contributions of the $\rho$ and $\omega$ intermediate states ($\sigma(s) - \sigma(s)|_{r_{\rho,\omega} = 0}$) and higher 
excitations ($\sigma(s) - \sigma(s)|_{A_{\phi', \rho', \omega'} = 0}$) are demonstrated in Fig.~\ref{crosspict} as an inset.

To study model dependence of the results, we perform alternative fits with the 
$A_{\phi',\rho',\omega'} $=0 amplitude in Eq.~(\ref{ksklxs}), or with an 
additional floating phase of the $\phi$ meson amplitude, or with the form of 
the inverse propogator $D_{V}(s) = m_{V}^{2} -s - i m_{V}\Gamma_{V}(s)$ instead 
of Eq.~(\ref{propogator}).
The variations of the $\phi$ meson parameters in these fits are used as an 
estimate of the model-dependent uncertainty presented as third errors in 
Table~\ref{endtable_phi_parameters}. 

As shown in Fig.~\ref{crossrel}, the obtained results have comparable accuracy 
but are not consistent, in general, with the previous data. 

The difference with the CMD-2~\cite{cmdc} measurement can be explained by the 
overestimation of the value of the trigger efficiency for slow kaons in the 
previous experiment. The positive  trigger decision of the CMD-2 required the 
presence of one charged track in DC in coincidence with the corresponding 
hits in the Z-chamber, and with at least one cluster in the CsI calorimeter 
with the energy deposition greater than 20 MeV.
But slow kaons stop in the first wall of the Z-chamber and only decay or 
their nuclear interaction products can make hits in the Z-chamber or leave 
energy in the calorimeter. The trigger efficiency about 90\% was obtained 
actually by simulation, using recorded information from detector cells.

In contrast to the CMD-2 experiment, the new CMD-3 detector has two 
independent trigger systems, the Z-chamber is excluded from the decision, and a 
charged (total) trigger efficiency is close to 100\%. The CMD-3 detector has 
the same Z-chamber and much more detailed information, and by including to 
our selection requirements of hits in the Z-chamber and presence of energy 
deposition greater than 20 MeV in the barrel calorimeter, we obtain 
a significantly larger trigger efficiency correction than the value obtained 
in the CMD-2 analysis~\cite{cmdc}. A reanalysis of CMD-2 data is expected.

Our value of $\Gamma_{\phi \to ee} B_{\phi \to  K^{+} K^{-}}$ is larger than the 
BaBar result by 1.8 standard deviations while the obtained value of 
$B_{\phi \to ee} B_{\phi \to K^{+} K^{-}}$ is larger than the PDG one, predominantly 
based on the CMD-2 measurement, by 2.7 standard deviations.
The obtained values of the  $\phi$ meson mass and width agree with the 
results of other experiments including our recent study of the process $e^+e^- \to K_{S}^{0}K_{L}^{0}$~\cite{ksklcmd3}.

\begin{table*}
\caption{The resulting parameters obtained from the cross section fit in comparison with previous experiments.}
\label{endtable_phi_parameters}
\begin{center}
\begin{tabular}{ccc}
\hline
                                Parameter                                          & CMD-3  & Other measurements          \\
\hline
$m_{\phi}$, MeV                                        &  1019.469 $\pm$ 0.006 $\pm$ 0.060 $\pm$ 0.010   & 1019.461 $\pm$ 0.019 (PDG2016) \\
$\Gamma_{\phi}$, MeV                                   & 4.249    $\pm$ 0.010 $\pm$ 0.005 $\pm$ 0.010  & 4.266 $\pm$ 0.031 (PDG2016)\\
$\Gamma_{\phi \to ee} B_{\phi \to  K^{+} K^{-}}$, keV  & 0.669   $\pm$ 0.001 $\pm$ 0.022 $\pm$ 0.005  & 0.634 $\pm$ 0.008  (BaBar)\\
$B_{\phi \to ee} B_{\phi \to K^{+} K^{-}}, 10^{-5}$    & 15.789  $\pm$ 0.033 $\pm$ 0.527 $\pm$ 0.120  & 14.24  $\pm$ 0.30 (PDG2016)\\
\hline
\hline
\end{tabular}
\end{center}
\end{table*}

\section{Contribution to $a_{\mu}$}
Using the result for the $e^+e^- \to K^{+} K^{-}$ cross section we compute the 
contribution of this channel to the anomalous magnetic moment of the muon 
$a_{\mu}$ via a dispersion relation in the energy region $2\cdot m_{K} < E_{\rm c.m.} < $ 1.06 GeV. 
According to Ref.~\cite{Davier_g_2}, for the leading-order 
approximation we obtain:

\begin{eqnarray}
\label{gm2formula}
a_{\mu}^{K^+K^-} =  \Bigr(\frac{\alpha m_{\mu}}{3\pi}\Bigr)^{2} \int_{4m_{K}^2}^{(1.06~\rm GeV)^2} \frac{ds}{s^2} K(s) \times \nonumber \\
 \times \frac{\sigma(e^+e^-\to K^+K^-)\cdot |1 - \Pi(s)|^2}{\sigma_0(e^+e^-\to \mu^+\mu^-)} = (19.33 \pm \nonumber \\ 
 \pm 0.04_{\rm  stat} \pm 0.40_{\rm  syst}  \pm 0.04_{\rm  VP}) \times 10^{-10}, ~~
\end{eqnarray}
where $K(s)$ is the kernel function, the factor $|1 - \Pi(s)|^2$ excludes 
the effect of leptonic and hadronic vacuum polarization (VP), and the Born cross 
section $\sigma_0(e^+e^-\to \mu^+\mu^-) = \frac{4\pi\alpha^2}{3s}$. The first 
uncertainty is statistical, the second one corresponds to the systematic 
uncertainty of $\sigma(e^+e^-\to K^+K^-)$ and the third is the uncertainty of the 
VP factor (0.2\%~\cite{Vacuumpolar}). 
We integrate Eq.~(\ref{gm2formula}) using the model for the cross section 
obtained in the previous section. Then, in order to avoid a model uncertainty, 
the difference between the experimental cross section values and used model 
is integrated using the  trapezoidal method.

The value should be compared with the recent result of the BaBar collaboration 
$a_{\mu}^{K^+K^-} = (18.64 \pm 0.16_{\rm stat} \pm 0.13_{\rm syst} \pm 0.03_{\rm VP}) \times 10^{-10}$ \cite{babarc} calculated in the same energy range.  
The difference between the calculation of $a_{\mu}$ based on our data and on 
the most precise previous measurement by BaBar is $1.6\sigma$.

\section{Comparison of $e^+e^- \to K^{+} K^{-}$ and $e^+e^- \to K_{S}^{0}K_{L}^{0}$ processes.}

\begin{figure*} 
   \begin{overpic}[scale=0.9]{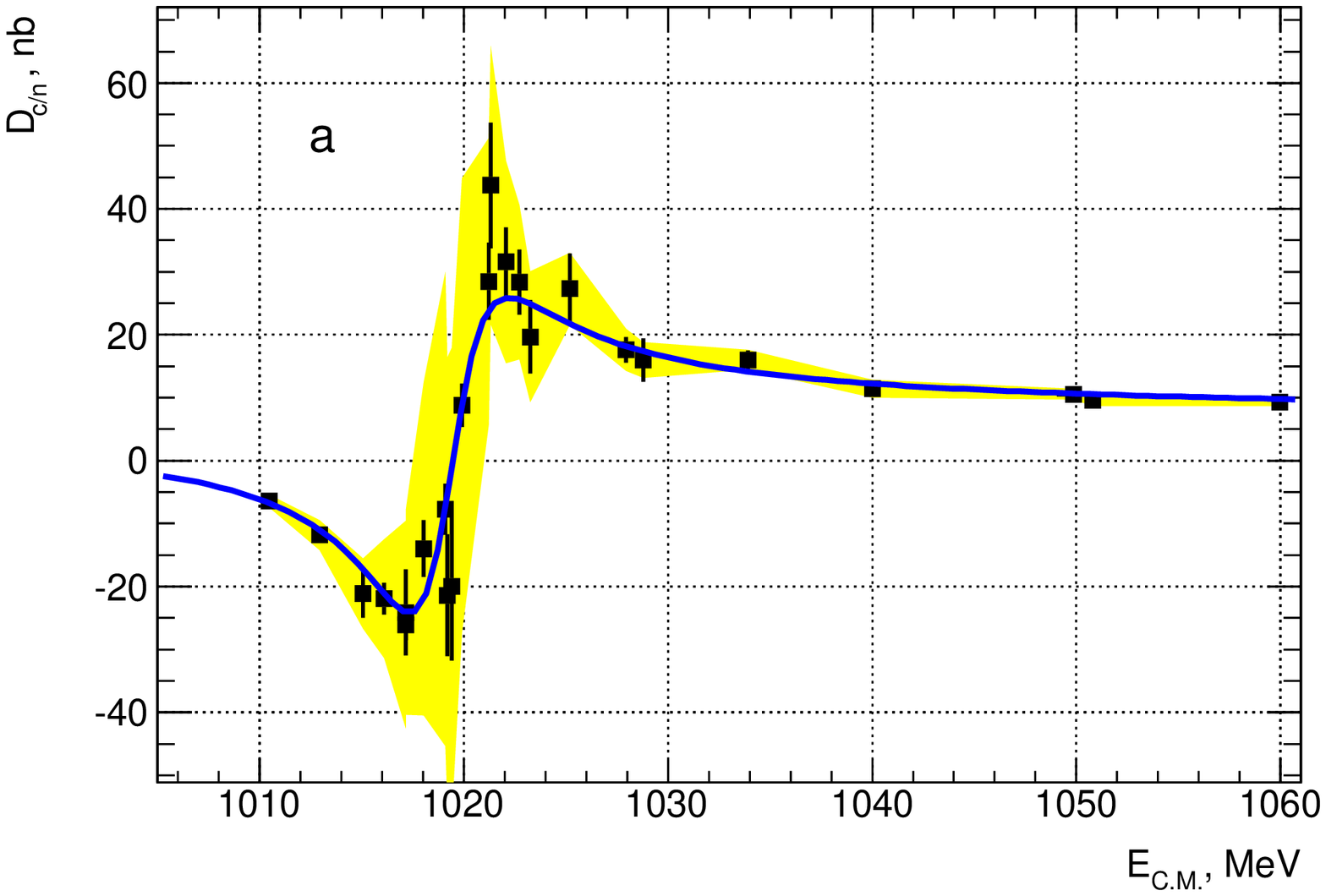}
 \put(60,41){\includegraphics[scale=0.3]{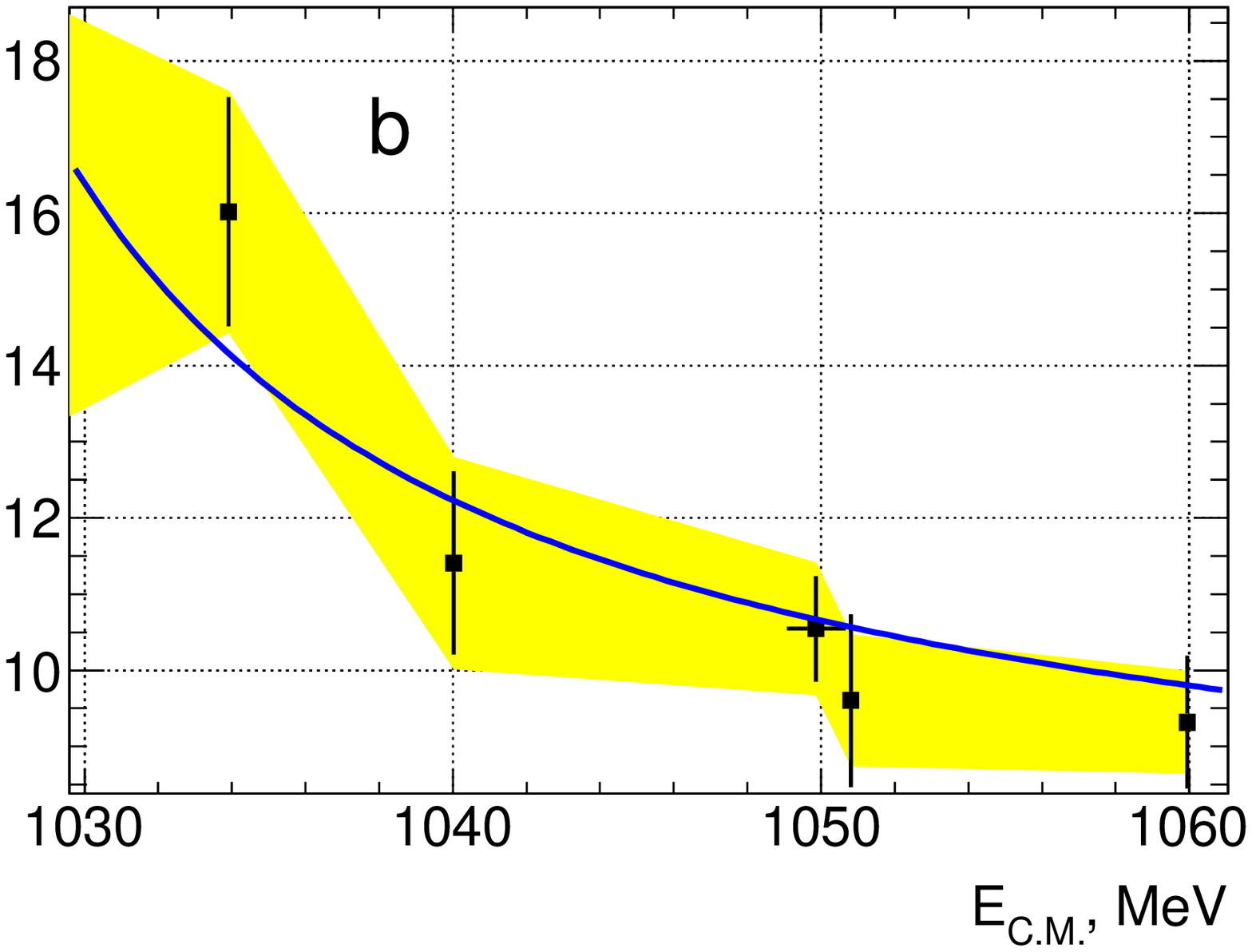}}
 \put(60,11.5){\includegraphics[scale=0.3]{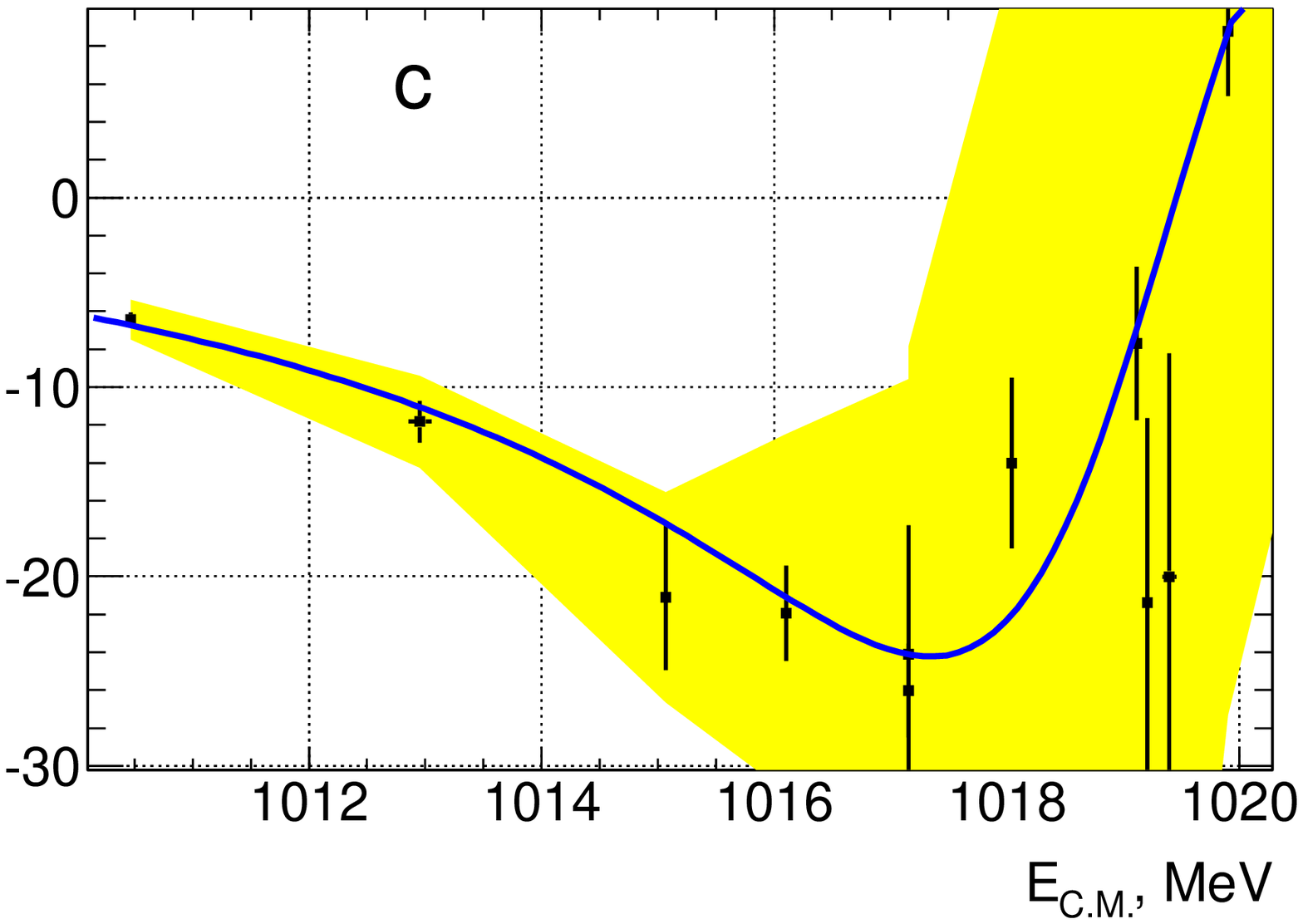}}
\end{overpic}
  \caption{ \label{cross_phi_rel}   
The difference of the charged and neutral cross sections defined as
$D_{c/n} = \sigma_{e^+e^- \to K^{+} K^{-}}\times \frac{\beta_{K^0}^3(s)}{\beta_{K^{\pm}}^3(s)} \times \frac{1}{Z(s)} - \delta_{K_{S}^{0}K_{L}^{0}} \times \sigma_{e^+ e^- \to K^0_{S}K^0_{L}}$.
The shaded area corresponds to systematic uncertainties in data, solid line - 
to the fit described in the text.
}
\end{figure*}

There is a strong relationship between the processes of electron-positron 
annihilation into $K^{+} K^{-}$ and $K^0_{S}K^0_{L}$ final states. The 
difference between them comes from the kinematic effect of the  $K^{\pm}$ 
and $K^0$ mass difference and  
the Coulomb interaction between $K^+$ and $K^-$ mesons~(\ref{FGSZ}).  At the 
$\phi$ peak the  Coulomb factor, $Z(m_{\phi}^2)$, contributes 4.2\% to the total 
cross section.
We correct the $\epem\to K^{+} K^{-}$ cross section for the above two 
effects and 
calculate the difference with the $\epem\to K_{S}^{0}K_{L}^{0}$ cross section: 
\begin{eqnarray}
\label{sfsdfsds}
D_{c/n} = \sigma_{e^+e^- \to K^{+} K^{-}}\times \frac{\beta_{K^0}^3(s)}{\beta_{K^{\pm}}^3(s)} \times \frac{1}{Z(s)} - \\ \nonumber
 - \delta_{K_{S}^{0}K_{L}^{0}} \times \sigma_{e^+ e^- \to K^0_{S}K^0_{L}},
\end{eqnarray}
where the factor $\delta_{K_{S}^{0}K_{L}^{0}}$ is introduced to account for a 
possible remaining systematic uncertainty in two measurements: most of the 
common uncertainties cancel in the difference.
The experimental value of $D_{c/n}$ is shown in Fig.~\ref{cross_phi_rel} by 
points with error bars, where the cross section of the production of neutral kaons  
is taken from our recent measurement~\cite{ksklcmd3}. The shaded area in the figure corresponds to the systematic 
uncertainties.

The deviation of $D_{c/n}$ from zero mostly comes from the different structure 
of the amplitudes of non-resonant isovector states, dominated by the $\rho$ 
meson, for the processes with charged and neutral kaons.
Indeed, instead of relations in Eq.~(\ref{sfdsfsdfsd}) for the charged final 
state the coupling constants of the $\omega(782)$ and $\rho(770)$ with the
$K^0_{S}K^0_{L}$ final state are:
\begin{eqnarray} 
\label{sfdasdasdassfsdfsd}
g_{\omega K^0_{S}K^0_{L}} = - g_{\rho K^0_{S}K^0_{L}} = - g_{\phi K^0_{S}K^0_{L}}/\sqrt{2},
\end{eqnarray}
where the $\rho$-meson term has a different sign. 
 So, the magnitude of $D_{c/n}$ in Eq.~(\ref{sfsdfsds}) is proportional to 
$\frac{g_{\rho KK} g_{\phi KK}}{D_{\phi}(s){D_{\rho}(s)}}$, that allows to see 
experimentally the $\rho$ meson contribution to K-meson production.

%More general: the isovector components in the  matrix elements of charged and neutral processes have opposite signs, and can be seen in the difference of Eq.~(\ref{sfsdfsds}). 

We fit $D_{c/n}$ using Eq.~(\ref{ksklxs},\ref{sfsdfsds}) with two floating 
parameters, $r_{\rho/\omega}$ and $\delta_{K_{S}^{0}K_{L}^{0}}$, discussed above. The mass, width of the $\phi$ meson and $\Gamma_{\phi \to ee} B_{\phi \to  K^{+} K^{-}}$ 
are fixed at the values obtained in Sec.~\ref{lapproxim}, 
also $\Gamma_{\phi \to ee} B_{\phi \to  K_{S}^{0}K_{L}^{0}}$ is fixed at 0.428 keV according to Ref.~\cite{ksklcmd3}.
The fit result is shown by a solid line in 
Fig.~\ref{cross_phi_rel}(a) and, in more detail, in insets to 
Fig.~\ref{cross_phi_rel} (b, c) and yields:
\begin{eqnarray} 
r_{\rho/\omega} = 0.954 \pm 0.027,  \nonumber \\
\delta_{K_{S}^{0}K_{L}^{0}} = 0.9964 \pm 0.0014, \nonumber \\
\chi^2/ndf = 22.2/22.  \nonumber 
\end{eqnarray}
We obtain good description of data by the fit. A small deviation of 
$r_{\rho/\omega}$ from unity demonstrates the precision ($\approx$ 5\%) of 
relations~(\ref{sfdsfsdfsd},\ref{sfdasdasdassfsdfsd}) and confirms that the
contribution from the $\rho(770)$ meson to $D_{c/n}$ dominates in the energy range under 
study. The deviation of $\delta_{K_{S}^{0}K_{L}^{0}}$ from unity (0.36\%) shows the 
level of a possible remaining systematic uncertainty of the cross section 
measurements.

Additionally, from the comparison of the charged and neutral cross sections we 
can obtain the ratio  of the coupling constants:                                
\begin{eqnarray}                                    
R = \frac{g_{\phi K^{+}K^{-}}}{g_{\phi K^0_{S}K^0_{L}} \sqrt{Z(m^2_{\phi})}}  ~~~~~~~~~~~~~~~~~~\nonumber \\
= \sqrt{\frac{B(\phi \to K^{+}K^{-})}{B(\phi \to K^0_{S}K^0_{L})}\cdot \frac{1}{Z(m^2_{\phi})}\cdot \frac{\beta_{K^0}^{3}}{\beta_{K^{\pm}}^3}} = 
  0.990 \pm 0.017, \nonumber
\end{eqnarray}
where the common parts of systematic uncertainties originating from those  
on the luminosity, radiative and energy spread corrections, are also reduced. 
As expected from isospin symmetry of u- and d-quarks, the value of $R$ is 
consistent with unity.

Additionally to the Coulomb interaction taken into account by the factor 
$Z(s)$, the final-state radiation of real photons, according to 
Ref.~\cite{Hoefer}, decreases the total $\epem\to K^{+}K^{-} (\gamma)$ 
cross section by about 0.4\% at the $\phi$ meson mass. This 
effect partially explains the deviation of $R$ from unity. 

\section{Conclusion}
Using  CMD-3 data in the \Ecm=1010--1060 MeV energy range we select 
1.7$\times$10$^6$ events of the process $e^+e^- \to K^{+} K^{-}$, and measure 
the cross section with an about 2\% systematic error. Using the fit in the 
VMD model the following values of the $\phi$ meson parameters have been 
obtained:
\begin{eqnarray} 
\mphi  \nonumber \\
\gammaphi     \nonumber \\
\gammaB   \nonumber 
\end{eqnarray}
We calculate the  contribution of the obtained cross section to the 
anomalous magnetic moment of the muon 
$\gmu $ in the energy range from 
threshold to $\sqrt{s}=$1.06 GeV.

The observed deviation of the $\rho(770)$ and $\omega(782)$ amplitudes, $r_{\rho/\omega} =   0.95\pm 0.03$, from a 
naive theoretical prediction  allows to 
estimate the precision of the used VMD-based phenomenological model as 
better than 5\%. The obtained ratio $\frac{g_{\phi K^{+}K^{-}}}{g_{\phi K^0_{S}K^0_{L}} \sqrt{Z(m^2_{\phi})}} =  0.990 \pm 0.017$ 
is consistent with isospin symmetry.

\section*{Acknowledgements}

We thank the VEPP-2000 personnel for the excellent machine operation. 
This work is supported in part by the Russian Education and Science Ministry, 
by the Russian Foundation for Basic Research grants RFBR 15-02-05674, 17-02-00897, 17-52-50064.

\end{document}

%% file: author_list.tex
\affiliation{Budker Institute of Nuclear Physics, SB RAS, 
Novosibirsk, 630090, Russia}
\affiliation{Novosibirsk State University, Novosibirsk, 630090, Russia}
\affiliation{Novosibirsk State Technical University, 
Novosibirsk, 630092, Russia}
\affiliation{University of Victoria, Victoria, British Columbia, Canada V8W 3P6}
\author{%
E.A.Kozyrev$^{1,2}$,
\quad E.P.Solodov$^{1,2}$,
\quad R.R.Akhmetshin$^{1}$,
\quad A.N.Amirkhanov$^{1,2}$,
\quad A.V.Anisenkov$^{1,2}$,
\quad V.M.Aulchenko$^{1,2}$,
\quad V.S.Banzarov$^{1}$,
\quad N.S.Bashtovoy$^{1}$,
\quad D.E.Berkaev$^{1,2}$,
\quad A.E.Bondar$^{1,2}$,
\quad A.V.Bragin$^{1}$,
\quad S.I.Eidelman$^{1,2}$,
\quad D.A.Epifanov$^{1,2}$,
\quad L.B.Epshteyn$^{1,2,3}$,
\quad A.L.Erofeev$^{1,2}$,
\quad G.V.Fedotovich$^{1,2}$,
\quad S.E.Gayazov$^{1,2}$,
\quad A.A.Grebenuk$^{1,2}$,
\quad S.S.Gribanov$^{1,2}$,
\quad D.N.Grigoriev$^{1,2,3}$,
\quad F.V.Ignatov$^{1}$,
\quad V.L.Ivanov$^{1,2}$,
\quad S.V.Karpov$^{1}$,
\quad A.S.Kasaev$^{1}$,
\quad V.F.Kazanin$^{1,2}$,
\quad A.A.Korobov$^{1,2}$,
\quad I.A.Koop $^{1}$,
\quad A.N.Kozyrev$^{1,2}$,
\quad P.P.Krokovny$^{1,2}$,
\quad A.E.Kuzmenko$^{1,2}$,
\quad A.S.Kuzmin$^{1,2}$,
\quad I.B.Logashenko$^{1,2}$,
\quad P.A.Lukin$^{1,2}$,
\quad A.P.Lysenko$^{1}$,
\quad K.Yu.Mikhailov$^{1,2}$,
\quad V.S.Okhapkin$^{1}$,
\quad E.A.Perevedentsev$^{1,2}$,
\quad Yu.N.Pestov$^{1}$,
\quad A.S.Popov$^{1,2}$,
\quad G.P.Razuvaev$^{1,2}$,
\quad Yu.A.Rogovsky$^{1,2}$,
\quad A.A.Ruban$^{1}$,
\quad N.M.Ryskulov$^{1}$,
\quad A.E.Ryzhenenkov$^{1,2}$,
\quad V.E.Shebalin$^{1,2}$,
\quad D.N.Shemyakin$^{1,2}$,
\quad B.A.Shwartz$^{1,2}$,
\quad D.B.Shwartz$^{1,2}$,
\quad A.L.Sibidanov$^{4}$,
\quad Yu.M.Shatunov$^{1}$,
\quad A.A.Talyshev$^{1,2}$,
\quad A.I.Vorobiov$^{1}$,
\quad Yu.V.Yudin$^{1,2}$
}